%% file: main.tex
\documentclass[conference, letterpaper]{IEEEtran}
\IEEEoverridecommandlockouts
\usepackage{cite}
\usepackage{graphicx}
\usepackage{url}
\def\BibTeX{{\rm B\kern-.05em{\sc i\kern-.025em b}\kern-.08em
    T\kern-.1667em\lower.7ex\hbox{E}\kern-.125emX}}
\makeatletter
\def\@IEEEsectpunct{.\ }
\renewcommand\paragraph{\@startsection{paragraph}{4}{\z@}%
  {0.5\baselineskip}{-0.5em}{\normalfont\normalsize\bfseries}}
\makeatother
\begin{document}

\title{SceneBaker: Radio-Ready Scene Generation\\ for Sionna Ray-Tracing
\thanks{
H. J. Yang is the corresponding author.}
}

\author{
    \IEEEauthorblockN{
        Hyeonsu Lyu\IEEEauthorrefmark{1},
        Minwoo Kim\IEEEauthorrefmark{3},
        Sojeong Park\IEEEauthorrefmark{3}
        and Hyun Jong Yang\IEEEauthorrefmark{1}\IEEEauthorrefmark{2}
    }
    \IEEEauthorblockA{
        \IEEEauthorrefmark{1} Institute of New Media and Communications, Seoul National University, Korea \\
        \IEEEauthorrefmark{2} Department of Electrical and Computer Engineering, Seoul National University, Korea \\
        \IEEEauthorrefmark{3} Department of Electrical Engineering, POSTECH, Korea \\
        \{hs.lyu, hjyang\}@snu.ac.kr, \{mwkim0210, sojeong\}@postech.ac.kr
    }
}

\IEEEaftertitletext{%
\begin{center}
\includegraphics[width=\textwidth]{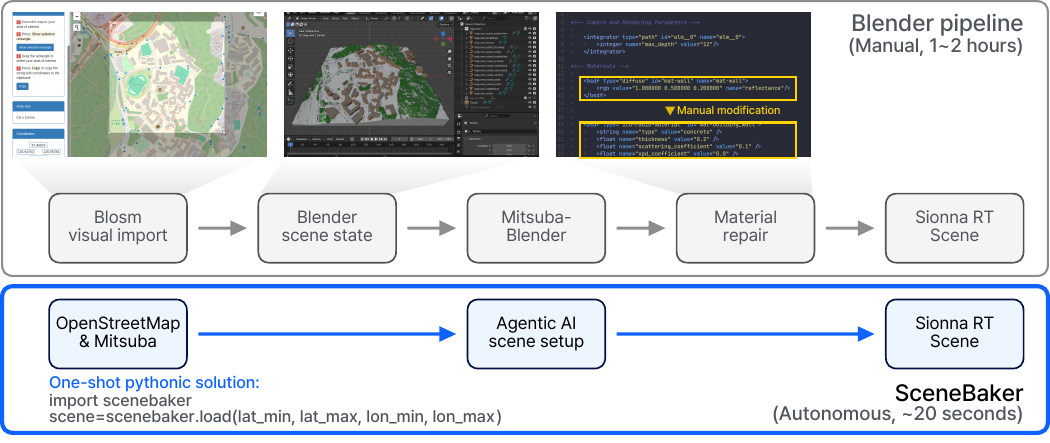}\\[3mm]
\refstepcounter{figure}\label{fig:pipeline_comparison}
\noindent\parbox{\textwidth}{\footnotesize Fig.~\thefigure. Existing and proposed scene generation pipelines over the same geographic bounds.}
\end{center}
\vspace{5mm}
}

\maketitle

\begin{abstract}
Wireless digital-twin (DT) research needs ray-tracing (RT) scenes that can be generated, versioned, and checked reproducibly.
Current visual-authoring workflows can produce plausible city models, but they are poorly matched to repeated radio-simulation studies because geometry, terrain contact, and material semantics often require manual repair after export.
This paper presents SceneBaker, a programmatic scene-generation pipeline that turns map and terrain data into Sionna RT-ready Mitsuba scenes without a GUI authoring step.
Across four campus scenes, SceneBaker generates Sionna RT-ready scenes over the same geographic bounds as a Blender-generated baseline.
The generated scenes avoid representative building-generation and terrain-contact faults while preserving comparable coverage-field and link-level channel-response behavior.
The implementation and generated comparison assets are available at \url{https://github.com/hslyu/sionna-scene-baker}.
\end{abstract}

\begin{IEEEkeywords}
Digital twin (DT), Sionna, ray-tracing (RT), OpenStreetMap, wireless propagation
\end{IEEEkeywords}

\input{sections/introduction}
\input{sections/pipelines}
\input{sections/evaluation}
\input{sections/conclusion}
\input{sections/references}

\end{document}

%% file: sections/introduction.tex
\section{Introduction}
Ray-tracing (RT)-based wireless digital twins (DTs) are becoming an important way to evaluate wireless deployments for next-generation wireless systems.
A DT maintains a software counterpart of a physical environment, so that the environment can be studied before it is changed \cite{minerva2020digital}.
By computing propagation from a modeled physical site, RT-based DTs enable site-specific wireless modeling beyond the abstractions of standardized stochastic channel models \cite{threegpp38901}.
This capability is expected to support next-generation wireless systems that require site-specific evaluation beyond conventional network-planning workflows \cite{khan2022dt6g,khan2022wirelessdt,bariah2023dtcommunications}.
In this context, Sionna RT has emerged as a practical platform that computes radio propagation paths from scene models and supports differentiable RT for environment-aware modeling \cite{hoydis2023sionna}.

An RT-based wireless DT requires a radio-ready scene that represents the physical site as reproducible simulation input.
The scene specifies the geometry, material assignments, and object placement from which the RT solver computes propagation.
Because this representation determines the RT results, scene construction must follow a consistent and reproducible procedure for RT-based wireless DTs to function as reproducible research platforms.

A practical bottleneck is that Sionna RT scene construction often starts from a handcrafted Blender workflow \cite{blender}.
Conventionally, this pipeline combines Blender with an OSM importer such as Blosm and the Mitsuba-Blender exporter \cite{sionnaRTIntro,blosmDocs}.
This pipeline is effective for inspecting a city model, but it makes the radio experiment depend on a GUI-centered authoring and export process.
Because this pipeline depends on third-party software components, its reproducibility is vulnerable to maintenance and compatibility changes across software versions \cite{blenderCompatibility,blenderAddons,mitsubaBlender}.
After export, the scene still requires radio-material repair because visual material definitions do not encode the propagation properties required by Sionna RT \cite{mitsubaBlender,sionnaRTMaterials}.
For wireless researchers, learning and repeating this pipeline can take roughly one to two hours per scene regeneration, which makes scene construction a bottleneck in an otherwise scriptable DT pipeline.

\begin{figure}[h]
\centering
\includegraphics[width=\columnwidth]{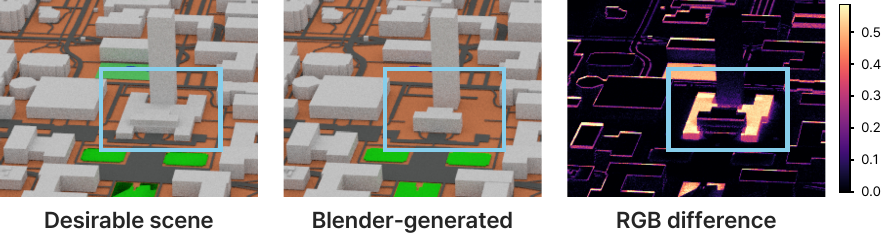}
\caption{Building-generation fault in the Blender-generated scene.}
\label{fig:ut_lobby_mismatch}
\end{figure}

This manual-authoring pipeline can also produce scenes that are inspectable but unreliable as RT inputs.
Figure~\ref{fig:ut_lobby_mismatch} shows a building-generation fault in a Blender-generated scene.
The example comes from the University of Texas at Austin campus, where a building present in the physical site is missing from the Blender-generated scene.
The resulting RT scene omits propagation-relevant building surfaces, which changes the geometry used by the RT solver.
The case is not important as a scene-specific detail; it represents a broader failure mode in which visual scene generation can fail to preserve physical structures that should remain part of the radio scene.

\begin{figure}[h]
\centering
\includegraphics[width=\columnwidth]{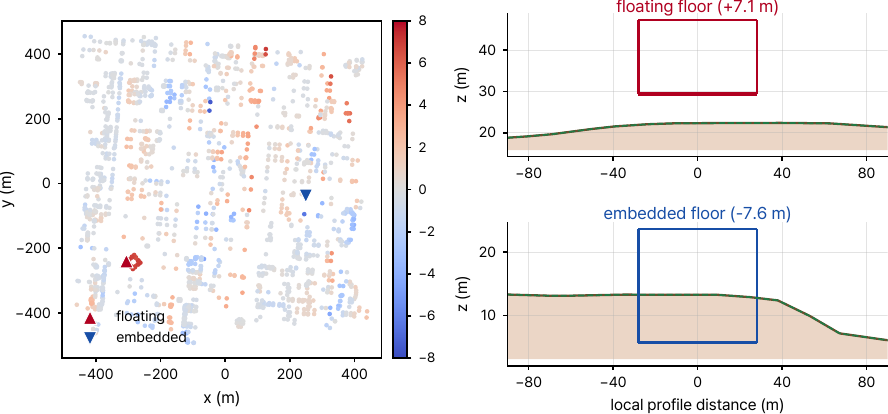}
\caption{Terrain-contact fault in the Blender-generated scene.}
\label{fig:terrain_contact_diagnostic}
\end{figure}

Terrain integration exposes the same limitation in a different form.
When a visual building model is placed independently from terrain, the ground can cut through the building or leave it floating.
Figure~\ref{fig:terrain_contact_diagnostic} shows this terrain-contact fault for the same campus area, where one building base is above the terrain while another is embedded below it.
The fault arises because building placement and terrain generation are not constrained by a shared geographic and vertical reference.
Such errors are not cosmetic in an RT scene because misplaced terrain contact changes which surfaces can block, reflect, or interact with propagation paths near the ground.

Radio-material repair creates another reproducibility problem as it makes material typing a user-dependent procedure.
Since the existing Blender pipeline requires manual radio-material assignment, users must decide which scene components should be mapped to propagation materials and which should be retained in the RT scene.
A single visual scene can become different RT inputs across users through different material choices or omitted propagation-relevant surfaces.
This makes it difficult to determine whether the difference in RT results comes from the method being evaluated or from the manual repair step.

This paper presents \emph{SceneBaker}, a programmatic pipeline for reproducible, autonomous generation of Sionna RT-ready scenes from geographic bounds without a handcrafted GUI-based workflow.
SceneBaker imports OSM-derived structure and optional terrain, performs agentic semantic scene setup for radio-material assignment, and writes the resulting scene directly for Sionna RT.
The contribution is a scene-construction workflow that automates the Blender-based pipeline while keeping geographic structure and semantic material ownership explicit within the same experimental procedure.
We evaluate SceneBaker on four campus scenes against Blender-generated baselines over the same geographic bounds.
The results show that SceneBaker avoids the representative faults above while producing near-equivalent scene footprints and comparable coverage-field and link-level channel-response behavior, supporting the claim that radio-ready scene generation can be made reproducible without relying on a manual Blender-centered pipeline.

%% file: sections/pipelines.tex
\section{Scene Construction Pipelines}
Scene construction for Sionna RT is not only a visual modeling problem, because the generated scene also defines the geometry, material ownership, and terrain placement used by the RT solver.
SceneBaker makes these propagation-relevant scene properties part of the generation procedure.
The Blender-centered workflow first produces a visual city model and then requires post-export repair before the scene can be used as a radio-simulation input.
Figure~\ref{fig:pipeline_comparison} compares these two workflows over the same geographic bounds.

\subsection{Existing Blender-centered pipeline}
The upper workflow in Fig.~\ref{fig:pipeline_comparison} follows the Blender-centered scene preparation pipeline described in the introductory Sionna RT workflow \cite{sionnaRTIntro}.
The user first prepares a compatible Blender environment with OSM-import and Mitsuba-export add-ons, then imports a selected OSM region into Blender as a visual city model.
The Mitsuba-Blender exporter converts this authored scene state into a Mitsuba scene for Sionna RT.
The scene must then be edited inside Blender to align the visual geometry with Sionna RT material and export conventions.
The simulation is completed in Python by loading the exported XML scene and placing the radio devices.
Because the exported scene inherits visual materials and authoring-time object organization, a post-export repair step is still needed to rewrite materials and group geometry for radio simulation.

This pipeline is useful as a baseline because it provides an inspectable city model.
Its limitation is that manual geometry grouping and radio-material repair can turn the same visual scene into different radio-simulation inputs.
Reproducing the scene requires preserving both the software environment and the user-dependent editing decisions used during scene construction.
This dependence weakens the objectivity and consistency required for comparing wireless-DT studies.

\subsection{SceneBaker pipeline}
SceneBaker converts geographic data directly into a Sionna RT scene, as shown in the lower workflow of Fig.~\ref{fig:pipeline_comparison}.
SceneBaker starts from a latitude-longitude bounding box, imports OpenStreetMap (OSM) structure through Overpass, and writes the generated scene as a Mitsuba XML scene with PLY geometry \cite{haklay2008osm,overpassAPI}.
The OSM nodes, ways, and relations are projected into a local metric coordinate system before they are converted into semantic mesh groups.
When terrain mode is enabled, the same conversion also incorporates Shuttle Radar Topography Mission (SRTM) elevation data as a standard terrain source \cite{farr2007srtm}.
The proposed framework directly generates Sionna RT-compatible scenes without relying on Blender or third-party Blender add-ons.

\paragraph{Semantic material ownership}
The semantic stage makes radio-material ownership a reproducible generation decision.
SceneBaker implements an agentic AI pipeline in which OpenAI's GPT-5.5 API interprets OSM-derived scene components and assigns them to radio-material classes before the Sionna RT-ready scene is written \cite{openaiGPT55}.
These assignments are preserved as mesh groups for built surfaces, circulation corridors, vegetated regions, water, open areas, and ground.
The XML writer then binds each group to a Sionna-compatible radio material, using ITU materials where a close physical category exists and custom radio materials otherwise.
SceneBaker records material identity through semantic mesh groups before export, making radio-material assignment an explicit scene-construction step.

\paragraph{Topology-derived building geometry}
The building stage generates the blocking structure that the RT solver uses.
SceneBaker reconstructs building footprints directly from closed OSM ways and multipolygon relations, separating them into distinct wall and roof meshes.
It then extracts vertical structure by processing available tags for height, level, base-height, and roof annotations.
Because height assignment is handled before export, missing OSM height metadata can be supplemented through retrieval-augmented generation (RAG) over external geospatial records before conservative defaults are applied \cite{lewis2020rag}.
This height-assignment step lets SceneBaker incorporate external height evidence while keeping building surfaces tied to geographic topology.

\paragraph{Surface and corridor geometry}
The surface stage defines the ground-level regions that can participate in propagation.
SceneBaker groups road and path centerlines by OSM highway class, expands them into finite-width surfaces, and resolves overlaps by applying a priority order among road groups.
Broad landuse and vegetation regions are clipped to the selected geographic bounds, and road cutouts are applied so that green areas do not cover traffic corridors.
SceneBaker derives surface layout from geographic semantics, so roads and land-cover regions remain explicit scene components.

\paragraph{Terrain-consistent placement}
The terrain stage gives the scene a common vertical reference.
When terrain mode is enabled, SceneBaker downloads the required SRTM HGT tiles, samples terrain height through the same local projection, and generates the ground as a gridded terrain mesh.
All generated surfaces are then placed by evaluating the same terrain model at their mesh vertices.
Polyline and polygon surfaces are subdivided when necessary so that long edges follow the terrain instead of floating across it.
SceneBaker addresses the terrain-contact failure shown in Fig.~\ref{fig:terrain_contact_diagnostic} by evaluating terrain height during scene generation and placing generated surfaces on the same vertical reference.

%% file: sections/evaluation.tex
\begin{figure*}[t]
  \centering
  \includegraphics[width=0.80\textwidth]{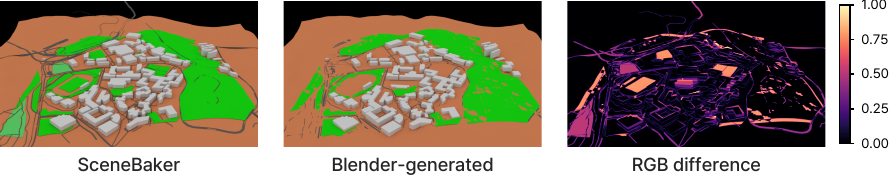}
  \caption{Semantic surface alignment under the same scene rendering view.}
  \label{fig:snu_terrain_render_comparison}
\end{figure*}

\section{Numerical Results}
The numerical study evaluates whether SceneBaker's automated scene generation preserves interpretable propagation behavior under matched geographic and simulation conditions.
The comparison is not a measurement-calibration study.
It evaluates consistency between two scene-generation pipelines under matched geographic bounds, transmitter placement, receiver configuration, solver settings, and material definitions wherever corresponding scene classes exist.
We organize the evaluation from scene representation to propagation behavior.
We first examine whether the generated scenes preserve campus-scale structure and semantic surface ownership, then test whether those surfaces are represented as solver-visible geometry, and finally compare the resulting coverage fields and link-level channel responses.

\subsection{Qualitative scene agreement}
We compare rendered terrain scenes to examine whether map-derived surface classes are spatially aligned with the generated radio scene.
This comparison matters because a wireless DT must attach user distributions and environment-dependent radio assumptions to scene surfaces with explicit semantic roles.
Roads, vegetation, and open areas carry operational meaning because they define where user density, mobility assumptions, and environment-dependent radio interactions can be placed.
Figure~\ref{fig:snu_terrain_render_comparison} compares the SNU terrain scene generated by SceneBaker with the Blender-generated baseline under the same rendering view.
Both pipelines preserve the main building layout and terrain envelope at the scale needed for a campus RT scene.
The important difference appears in ground-cover representation.
In the Blender-generated scene, several road and vegetation regions remain fragmented or weakly tied to stable semantic surface ownership.
SceneBaker reconstructs these regions as explicit radio surfaces with material ownership, making the generated scene more suitable for later DT-level mapping of users and environmental assumptions.
The SNU render difference has a median RGB norm of \(0.004\), while \(18.1\%\) of pixels exceed a norm difference of \(0.1\).
The high-difference pixels are concentrated in surface classes, while the main building mass remains largely aligned.
The RGB difference panel is read here as evidence of semantic surface reconstruction, not as a pixel-level error against Blender.
This qualitative result motivates the structural analysis below, because semantic surface ownership is useful only when it becomes part of the geometry visible to the propagation solver.

\subsection{Scene-level structure}
\begin{figure}[t]
\centering
\includegraphics[width=\columnwidth]{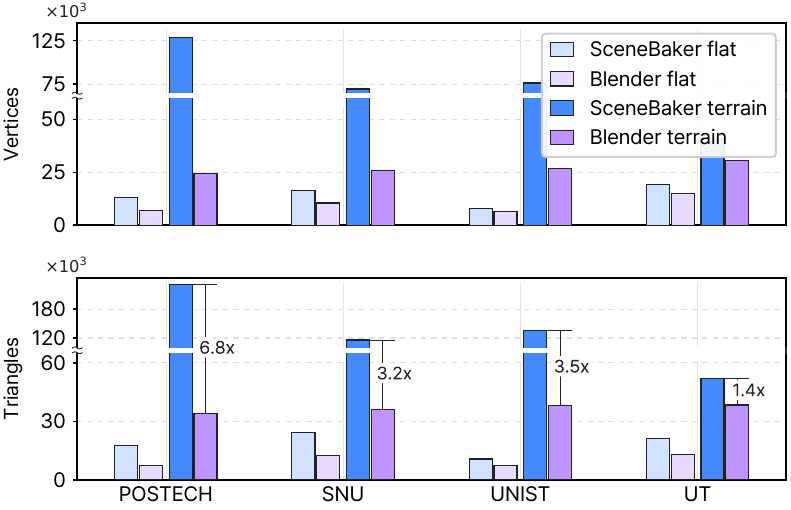}
\caption{Solver-visible mesh complexity of generated radio scenes.}
\label{fig:radio_scene_mesh_complexity}
\end{figure}

We next compare mesh complexity to determine whether the semantic surfaces identified above are represented as solver-visible geometry.
Figure~\ref{fig:radio_scene_mesh_complexity} summarizes vertex and triangular-face counts for the flat and terrain versions of each campus scene generated by SceneBaker and by the Blender workflow.
This comparison is needed because propagation gaps are difficult to interpret if the added geometry could originate from arbitrary changes in the main building mass.
The flat variants provide a footprint-level reference, while the terrain variants expose the additional geometric cost of representing ground-contact surfaces and semantic land-cover classes.
We also interpret building geometry separately from terrain-attached surfaces, because uncontrolled building changes would weaken the later propagation comparison.
The resulting pattern shows that SceneBaker's additional resolution is concentrated in terrain-attached semantic surfaces.
The building mass remains comparatively stable across the two pipelines.
This pattern indicates that SceneBaker adds mesh resolution to represent geographically meaningful surfaces as solver-visible geometry.
These surfaces can support both propagation analysis and future DT-level user-distribution mapping.
The following coverage-field comparison evaluates the radio effect of a scene representation whose added complexity has an interpretable semantic source.

\subsection{Coverage-field agreement}
\begin{figure}[!t]
\centering
\includegraphics[width=\columnwidth]{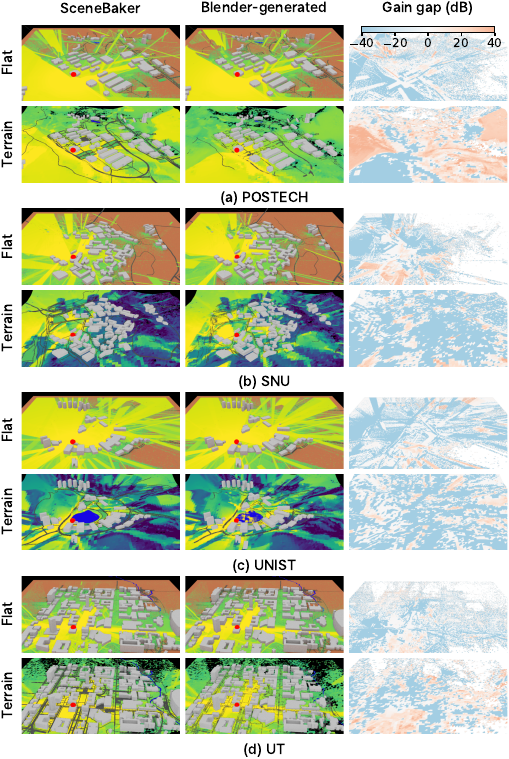}
\caption{Coverage-field comparison between SceneBaker and Blender-generated radio scenes.}
\label{fig:radiomap_solver_main}
\end{figure}

We compare coverage fields to evaluate whether SceneBaker preserves the spatial path-gain behavior of the Blender-generated baseline.
Figure~\ref{fig:radiomap_solver_main} compares the SceneBaker and Blender-generated path-gain fields across the campus scenes.
This evaluation is needed because structural agreement alone does not show whether the generated scene produces comparable radio behavior under the same transmitter and receiver configuration.
All cases use matched geographic bounds, transmitter placement, receiver configuration, material definitions, and solver settings.
This control makes the reported gap reflect scene generation, not simulation configuration.
We quantify the gain gap shown in the difference panels as
\(\Delta G = 10\log_{10}G_{\mathrm{SB}} - 10\log_{10}G_{\mathrm{BL}}\)
over cells where both pipelines produce finite positive path gain.
Across \(1.26\times 10^{7}\) valid cells, the pooled median absolute gain gap is \(0.276\) dB and the 90th-percentile absolute gap is \(8.14\) dB.
At the case level, flat scenes have a median absolute gain gap of \(0.303\) dB, indicating that the footprint-level building and surface layout preserves the coverage field when terrain interaction is removed.
Terrain scenes are stricter because ground contact, elevation, and semantic surface reconstruction become part of the propagation geometry; their case-level median absolute gap increases to \(4.25\) dB.
The largest case-level median is \(9.75\) dB in the POSTECH terrain scene, where local terrain-sensitive interactions dominate the comparison.
The larger terrain gaps are consistent with the structural result above, because terrain-attached semantic surfaces change the geometry that the RT solver uses for ground interaction.
The coverage-field result supports consistency within a shared propagation regime, not calibrated equivalence to measured propagation.

\begin{figure}[t]
\centering
\includegraphics[width=\columnwidth]{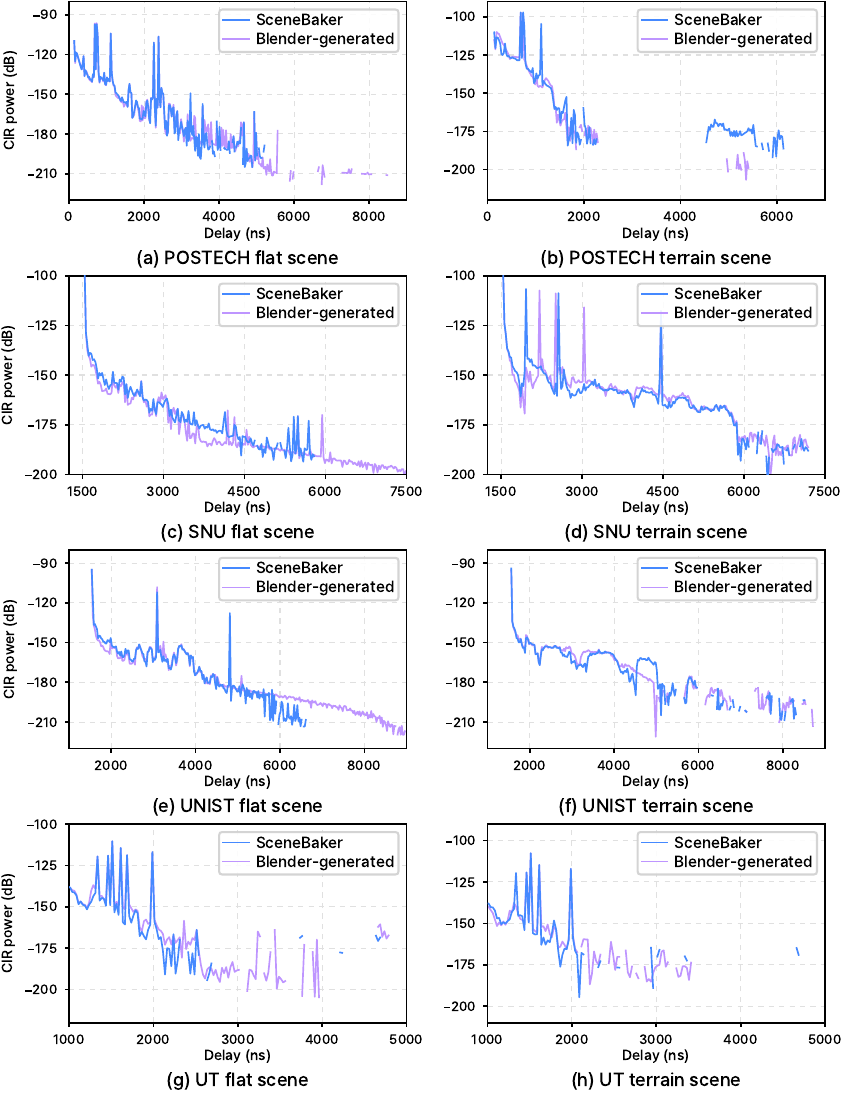}
\caption{Link-level channel impulse response comparison between generated radio scenes.}
\label{fig:pathsolver_cir_main}
\end{figure}

\subsection{Link-level channel-response agreement}
We compare link-level channel responses to test whether the coverage-field agreement persists under a more sensitive delay-domain metric.
Figure~\ref{fig:pathsolver_cir_main} compares the binned channel impulse response (CIR) power curves for the same transmitter and receiver placement.
This comparison is needed because coverage fields aggregate propagation behavior spatially, whereas individual links can expose changes in dominant paths, blockage, and reflection timing.
Using the 25-ns delay bins used in the plot, the pooled median absolute CIR-power gap over common nonempty bins is \(2.02\) dB, and the 90th-percentile absolute gap is \(10.75\) dB.
The case-level median absolute CIR gap ranges from \(0.65\) to \(4.17\) dB, which is consistent with the coverage-field result that most scene differences remain moderate after aggregation.
The total received path-gain difference has a median absolute value of \(0.41\) dB across the eight cases, showing that the aggregate received power is usually preserved even when individual paths are reorganized by terrain and surface geometry.
The UT terrain case is the main exception: it reaches \(8.78\) dB in total gain difference and changes the strongest component by \(13.95\) dB.
This outlier is consistent with the limitation identified in the Introduction, because terrain contact and local blocking surfaces can change a small number of dominant paths even when the broader scene footprint remains aligned.
The link-level result complements the coverage-field evaluation by showing that most aggregate behavior is preserved, while a small number of dominant components can change when terrain contact and surface geometry differ.
Together, these results indicate that SceneBaker preserves the main propagation behavior of the Blender-generated baseline while making local geometry differences visible in delay-domain responses.

%% file: sections/conclusion.tex
\section{Conclusion}
This paper presented SceneBaker as a radio-scene generation pipeline for Sionna RT.
The motivating problem is that visual city-authoring tools can produce plausible scenes while still requiring manual repair of materials, topology, and terrain contact before radio simulation.
SceneBaker provides a direct geographic-to-radio scene-generation path from geographic bounds, OSM semantics, and optional terrain data to Sionna RT scenes.
Across four campus scenes, this approach avoids representative faults in the Blender-generated baseline and produces comparable scene-level and solver-level outputs.
These results make SceneBaker suitable as a research platform for generating consistent radio scenes with explicit assumptions.
At the same time, consistency is not the same as fine-grained real-world fidelity, and generating scenes that closely match the physical environment remains an open problem.
The current implementation cannot recover building information that is absent from OSM.
Missing building-height evidence is handled by conservative defaults when no external evidence is retrieved.
The material parameters used here are initial values, and they have not been calibrated against site measurements.
SRTM terrain also cannot represent small local ground discontinuities.
Future work will calibrate material parameters against measurements, incorporate richer building evidence, and evaluate propagation agreement over measured campus links.

%% file: sections/references.tex
\bibliographystyle{IEEEtran}
\bibliography{references}